\documentstyle[preprint,aps]{revtex}
\date{\today}
\begin{document}
\draft
\title{\bf Incommensurate Phase on a Disordered Surface:
Instability Against the Formation of Overhangs and Finite Loops}
\author{K. Ziegler}
\address{Institut f\"ur Theorie der Kondensierten Materie,
Universit\"at Karlsruhe, Physikhochhaus, D-76128 Karlsruhe, Germany}
\author{A.M.M. Pruisken}
\address{Instituut voor Theoretische Fysica, Universiteit van Amsterdam,
NL-1018 XE Amsterdam, The Netherlands}
\maketitle
\begin{abstract}
The stability of the quenched incommensurate phase in two dimensions against
the creation of overhangs and finite loops (OH/FL) in the replica space is
investigated for a model of domain walls with $N$ colors.
Introducing a chemical potential $\epsilon$ for OH/FL,
the probability for the formation of these objects is studied for
$\epsilon\to0$. In the pure limit this probability vanishes with
$\sqrt{\epsilon}$, whereas the fluctuations of this probability are
long-range correlated in the quenched system. This indicates an instability
related to the symmetry in replica space. It is accompanied by the creation
of a massless boson. The latter leads to a power law decay with exponent
$\propto 1/N$ for the product of the correlation functions along the domain
walls.
\end{abstract}
The model for the commensurate-incommensurate transition (CIT) in $d=1+1$
dimensions is the prototype for a class of systems
characterized by directed domain walls, directed random walks, directed
polymers or flux lines.
The common feature is that of interacting random walks where the walkers
choose randomly steps forward, left or right but not backward. The
interaction is due to the condition that walks are not allowed to cross.
In the presence of quenched disorder these systems should exhibit a new
phase which is related to a glass or frozen phase. The latter
can be regarded as freezing of the domain walls (directed random walks
etc.) in a random potential. From a conceptional point of view the CIT
model has attracted attention because
it is soluble without disorder\cite{1}, and the effect of disorder
can be regarded as a perturbation.
The disordered case has been studied by various methods
like the Bethe ansatz\cite{2}, scaling arguments\cite{3}, variational
approaches\cite{4} and with perturbation theory\cite{5}. Although these
works agree on the fact that the critical exponent of the density changes
from $1/2$ in the pure system to $1$ in the disordered system they came to
different conclusions concerning the decay of the
correlation function in the quenched system.
In a recent article by Tsvelik\cite{6} it was argued that quenched disorder
is irrelevant with respect to the asymptotic decay of the density density
function. This
is surprising since the correlation function should be more sensitive to
additional fluctuations than the density. Therefore, at least near the CIT
one should expect that disorder changes the qualitative properties of the
model. For the corresponding model with disorder correlated along
the direction of the domain walls was found that the long-range correlation
is destroyed\cite{7}. In the following we will present a calculation which
shows that there is indeed a new feature due to disorder which affects the
correlations. It is characterized by the formation of overhangs and finite
loops (OH/FL).
A crucial point in the various approaches to the CIT in the presence of
disorder is the replica trick. It has been argued that there is no
replica symmetry breaking (however, see \cite{8}). In contrast to this
we will start from a broken replica symmetry given by an external field
(chemical potential) which creates OH/FL.

The grand canonical statistics of domain walls in $d=1+1$ dimensions can be
described by
a fermion Lagrangian density in imaginary time representation\cite{6}
\begin{equation}
L=c^+\partial_\tau c-\partial_xc^+\partial_xc-v(x,\tau)c^+c+\mu c^+c
\end{equation}
and the partition function
\begin{equation}
Z=\int{\cal D}[c^+,c]\exp(-\int_0^\infty d\tau\int dx L).\end{equation}
The imaginary time $\tau$ is along the direction of the domain
walls and $x$ is perpendicular to $\tau$.
Disorder is introduced by the random potential $v(x,\tau)$ which is Gaussian
distributed with zero mean and
$\langle v(x,\tau)v(x',\tau')\rangle=\sigma^2\delta(x-x')\delta(\tau-\tau')$.

The partition function is invariant under a time-reversal transformation.
Only the time differential operator is changed as
$\partial_\tau\to\partial_\tau^T=-\partial_\tau$ by this transformation. The
integration of the non-interacting fermions in $Z$ gives the space-time
determinant $\det(\partial_\tau+\partial_x^2+\mu-v)$. Then the time-reversal
invariance of $Z$ is obvious because the transposition of the matrix leaves
the determinant invariant.

Averaging of the free energy or of an expectation value
can be performed either by using a supersymmetric
generalization of the fermion Lagrangian\cite{9} or by
using the replica trick $\ln Z=\lim_{n\to0}(Z^n-1)/n$. Since both
representations are equivalent if we choose the correct
structure, we will use the replica trick here for convenience. Introducing
an even number of replicas, say $2n$, we take one
sector of $n$ replicas with $\partial_\tau$
and the other sector of $n$ replicas with $\partial^T_\tau$.
This can be written in a spinor representation for the Lagrangian density
as
\begin{equation}
L=\pmatrix{
c^+\cr
d^+\cr
}\cdot\pmatrix{
0&\partial_\tau+\partial^2_x-v+\mu\cr
\partial^T_\tau+\partial^2_x-v+\mu&0\cr
}\pmatrix{
d\cr
c\cr
},\label{la2}
\end{equation}
where $c$, $d$ are $n$-component fermions, and the matrix in (\ref{la2})
is diagonal w.r.t. to the $n$ replica components.
Now we introduce a chemical potential $\epsilon$ which allows the formation
of OH/FL by combining the two different replica sectors. This means that
the chemical potential creates a particle-hole pair $c^+d$ which
can be annihilated at another space-time point by $d^+c$. Including
this new chemical potential in the Lagrangian density (\ref{la2}) we
replace the zeros in the diagonal elements by $i\epsilon$.
The imaginary unit $i$ guarantees a positive weight for the OH/FL
in the partition function because OH/FL contain
always a sequence of reversed Grassmann variables which contribute a
minus sign. The density of OH/FL can be measured by varying locally the
chemical
potential: $\rho_\epsilon(x,\tau)\propto\langle c^+(x,\tau)d(x,\tau)+
d^+(x,\tau) c(x,\tau)\rangle$.
In the pure limit $v=0$ this density can be evaluated by a simple
calculation as $\rho_\epsilon\sim const.\sqrt{\epsilon}$; i.e.,
it vanishes continuously for vanishing
chemical potential $\epsilon$. Thus, the pure system is {\it stable} against
the formation of OH/FL. Another interesting quantity is related to the fact
that OH/FL are given by a combination of local edges (s. Fig.1)
which are separated by a distance in $x$ and $\tau$. The correlation of
an edge at $(x,\tau)$ and another one at $(0,0)$ is
\begin{equation}
\langle c^+(x,\tau)d(x,\tau) d^+(0,0)c(0,0)\rangle\label{edcor}.
\end{equation}
Because this is a correlation function of non-interacting fermions it will
factorize into $\langle c^+(x,\tau)d(x,\tau)\rangle\langle d^+(0,0)c(0,0)
\rangle -\langle c^+(x,\tau)c(0,0)\rangle\langle d^+(0,0)d(x,\tau)\rangle$.
The first term measures the density of OH/FL whereas the second term is
the product of correlations along the domain walls from $(0,0)$ to
$(x,\tau)$.

Our calculation will be based on the generalization of the domain wall model
to one which has walls with $N$ different colors. Without quenched disorder
the new model separates into $N$ independent models. However, the
disorder potential is chosen as $v_{\alpha\beta}(x,\tau)$ with $\alpha,
\beta=1,...,N$ such that the domain walls may change statistically the
color from $\alpha$ to $\beta$. This mechanism reduces the hard-core of the
domain walls (i.e., Pauli's principle of the fermions) because the
domain walls can avoid each other by changing the color.
The complex matrix elements $v_{\alpha\beta}$ are statistically
independent except for the symmetry $v_{\alpha\beta}=v_{\beta\alpha}^*$.
After averaging we obtain an effective interaction for the fermions
(${\hat c}=(d,c)$) in the replica model
\begin{equation}
{\sigma^2\over2N}\sum_{i,j=1}^{2n}\sum_{\alpha,\beta=1}^N{\hat c}^+_{i,\alpha}
{\hat c}_{i,\beta}{\hat c}^+_{j,\beta}{\hat c}_{j,\alpha}.
\end{equation}
The limit $N\to\infty$ can be solved by a saddle approximation. This becomes
obvious if one decouples the effective fermion-fermion interaction of the
quenched model by a new random matrix field $Q$ which includes fluctuations
of both chemical potentials $\mu$ and $\epsilon$: we replace $\sigma^2N^{-1}
{\hat c}^+_{i,\alpha}{\hat c}_{i,\beta}{\hat c}^+_{j,\beta}
{\hat c}_{j,\alpha}=-\sigma^2N^{-1}{\hat c}^+_{i,\alpha}{\hat c}_{j,\alpha}
{\hat c}^+_{j,\beta}{\hat c}_{i,\beta}$ by $2iQ_{ji}\sum_\alpha
{\hat c}^+_{i,\alpha}{\hat c}_{j,\alpha}$. $Q$ is a Hermitean $2n\times2n$
matrix field. Going
back to the replicated partition function $Z^{2n}$ we integrate out the
fermion field ${\hat c}_{i,\alpha}$. This leads to an effective action which
depends only on the decoupling field as
\[
S_{eff}=
N\int_0^\infty d\tau\int dx{2\over\sigma^2}Tr_{2n}(Q(x,\tau)^2)
\]
\begin{equation}
-N\ln{\rm det}\pmatrix{i\epsilon+2iQ_{21}(x,\tau)&\partial_\tau+\partial^2_x+
\mu+2iQ_{11}(x,\tau)\cr
\partial_\tau^T+\partial^2_x+\mu+2iQ_{22}(x,\tau)&i\epsilon+2iQ_{12}(x,\tau)
\cr
}.\end{equation}
$Tr_{2n}$ denotes the trace w.r.t. the $2n$ replica and ${\rm det}$
the determinant w.r.t. the $2n$ replica {\sl and} $x,\tau$.
This is the 1+1-dimensional replica version of the $2+1$-dimensional
`supersymmetric' action found for the flux lines in a random potential
\cite{9}.

$S_{eff}$ depends on the number of colors only through the prefactor $N$.
Therefore, the effective field theory for large $N$
enables us to do the functional integration in saddle point
approximation. This approximation can be interpreted as the
replacement of the chemical potentials $\mu$ and $\epsilon$ in the free
fermion propagator by a self-energy term
which is determined by the saddle point equation $\delta S_{eff}=0$.
There are two contributions if we assume that $Q$ provides only
additive corrections to the two chemical potentials. One is a shift of the
chemical potential $\mu$ in the limit $\epsilon\to0$
\begin{equation}
\mu_s\equiv 2iQ_{jj}=-\sigma^2\int dkd\omega{-k^2+\mu+\mu_s\over
D(\epsilon')}{\ \ \ \ }(j=1,...,2n)
\label{mu}
\end{equation}
with  $D(\epsilon')=\epsilon'^2+\omega^2+(-k^2+\mu+\mu_s)^2$,
and the other is a shift of the chemical potential for OH/FL
\begin{equation}
\epsilon'\equiv 2Q_{jj+n}=2Q_{j+nj}=\epsilon'\sigma^2\int dkd\omega/
D(\epsilon'){\ \ \ \ }(j=1,...,n).\label{epsilon'}
\end{equation}
$\epsilon'=0$ is always a solution of (\ref{epsilon'}). This is a
replica-symmetric (RS) solution. A replica-symmetry breaking (RSB) solution
$\epsilon'\ne0$ can be found if
\begin{equation}
\int dkd\omega/D(0)>1/\sigma^2\label{rsb}.\end{equation}
Since the denominator $D(\epsilon')$ is increasing with increasing
$\epsilon'^2$ there is an $\epsilon'$ which satisfies
\begin{equation}
\int dkd\omega/D(\epsilon')=1/\sigma^2.\end{equation}
This can be rewritten as
\begin{equation}
\epsilon'^{1/2}=\sigma^2\int d{\bar k} d{\bar\omega}
{1\over 1+{\bar\omega}^2+(-{\bar k}^2+\mu'/\epsilon')^2}
.\end{equation}
In this case we obtain from (\ref{mu}) a renormalized chemical potential
\begin{equation}
\mu'\equiv\mu+\mu_s
=\mu-\epsilon'^{1/2}\sigma^2\int d{\bar k} d{\bar\omega}{-{\bar k}^2+
\mu'/\epsilon'\over 1+{\bar\omega}^2+(-{\bar k}^2+\mu'/\epsilon')^2}.
\end{equation}
$D(\epsilon')$ is increasing with decreasing $\mu'<0$. This implies that
there is a $\mu_c<0$ (which depends on $\sigma$) such that
$\int dkd\omega /D(0)<1/\sigma^2$ for $\mu'<\mu_c$. Consequently,
there is only a RS solution $\epsilon'=0$ for $\mu'<\mu_c$. This behavior
describes a transition from a RS solution to a RSB solution if we change
$\mu$ or the disorder $\sigma$.

The evaluation of the density of domain walls in saddle point approximation
requires a cut-off, because only a finite number of domain
walls per unit length is reasonable. With the cut-off $k^2=1$ we obtain
after integration over $\omega$
\begin{equation}
\rho\propto1-{i\over \sigma^2}(Q_{11}+Q_{22})=1+\int_0^1dk{-k^2+\mu'\over
\sqrt{\epsilon'^2+(-k^2+\mu')^2}}.\end{equation}
$\mu'<0$ and $\epsilon'\to 0$ implies $\rho\to0$. The density
vanishes linearly. For instance, the density of the RS solution is $\rho
\propto\sigma^2+\mu$. Thus the density of the RS as well as the RSB
solution are in agreement with the linear behavior of the Bethe ansatz
calculation of Kardar\cite{2}.

Finally we evaluate the Gaussian fluctuations around the saddle point solution
which are the $1/N$ corrections. The stability matrix can be taken from
Ref. \cite{9}. The fluctuations of $Q_{11}$, $Q_{22}$
are massive with the eigenvalues $\lambda_\pm$
\begin{equation}
\lambda_+=\lambda_-+4\int dkd\omega\omega^2/ D(\epsilon')^2.
\end{equation}
For $\epsilon'\ne0$ $\lambda_-$ is
\begin{equation}
\lambda_-=4\epsilon'^2\int dkd\omega/D(\epsilon')^2;
\end{equation}
i.e., the eigenvalues of the RSB solution are always positive (stable).
On the other hand, for $\epsilon'=0$ the eigenvalue $\lambda_-$ is
\begin{equation}
\lambda_-=2/\sigma^2-2\int dkd\omega/D(0).\label{ev}
\end{equation}
Since inequality (\ref{rsb}) holds if there are two saddle point
solutions, $\lambda_-$ of the RS solution (\ref{ev})
is negative (unstable) when the RSB solution exists.
The other two eigenvalues of the RSB solution are $\lambda_3=2(\lambda_+
-\lambda_-)>0$
(related to $Q_{12}-Q_{21}$) and the massless one $\lambda_4=0$
(related to $Q_{12}+Q_{21}$). The massless fluctuations are a consequence of
a global symmetry of the model \cite{9}. They can be expressed as a
non-linear sigma model for $Q$
with the constraint
${\bf Q}^2={\bf 1}$, $Tr_{2n}{\bf Q}=0$ and ${\bf Q}\sigma_2{\bf Q}=
-\sigma_2$. The constraint can be satisfied if
we parametrize ${\bf Q}$ by an $n$-component field $\varphi$ as
$Q=\sigma_3(\cos^2\varphi-\sin^2\varphi)+2\sigma_1\cos\varphi
\sin\varphi$. This parametrization neglects a unitary transformation inside
of each replica sector which is not expected to affect the
properties of the $\varphi$-dependent part of the model. With the field
$\varphi$ the action of the non-linear sigma model reads
\begin{equation}
S_{eff}=bN\int dxd\tau\varphi(\partial_x^2+\partial_\tau^2)\varphi
\end{equation}
with a positive constant $b$, depending on $\gamma$ but not on $N$
or $\epsilon$,
which are given in Ref.\cite{9}. From this bilinear action we can evaluate
the correlation function of the density of the edges:
\begin{equation}
\langle Q_{12}(x,\tau)Q_{21}(0,0)\rangle\sim{\epsilon'}^2const.\Big(x^2+\tau^2
\Big)^{-1/2\pi bN}\label{corrf}
\end{equation}
for large distances. Thus there is no replica symmetry breaking because the
correlation of the order parameter decays.
According to our previous remark this implies also a power law decay for the
product of correlation functions along the domain walls.
In contrast, a single averaged
correlation function decays exponentially according to the finite length
scale created by ${\epsilon'}^{-1}$. On a pure surface the correlation
function decays like $(x^2+\tau^2)^{-1/2}$.

In conclusion, we have found that the statistics of domain walls in the
CIT model with quenched disorder is unstable against the formation of
OH/FL in the replica space. This result was based
on a saddle point calculation for an effective self-energy in the
replica model with $2n$ replica and $N$ colors. The properties of
the model are similar to those of the Gross Neveu model\cite{12} and other
1+1-dimensional fermion models with attractive interaction\cite{13}.
The creation of OH/FL in the quenched CIT model is plausible because the
random potential favors OH/FL which can easily circumvent
an unfavorable potential or freeze into the local potential
wells. Since there are no OH/FL by definition,
the replica model creates them spontaneously by combining wall elements
coming from different replica sectors. However, the effect of the OH/FL is
only marginal in our calculation because they appear with vanishing density.
Nevertheless, the instability the instability related to the OH/FL means
a drastic change in the quenched state compared with the pure CIT problem.

A similar instability was also discussed in the context of
the Gross Neveu model by Witten\cite{14} who evaluated the correlation
function of the massless bosons. Those bosons also lead to a power law decay
with an exponent proportional to $1/N$.

It should be emphasized that the formation of OH/FL is a special type of
instability regarding the freezing into local minima
of the random potential. We cannot exclude other instabilities which
may be even more favorable. Therefore, the purpose of this investigation was
only to provide an example for an instability in the quenched CIT
model. It indicates that the model has probably a rich structure which
deserves further investigations based on effects related to replica
symmetry.

\begin{figure}
\caption{A finite loop (a) and an overhang (b) created for the domain wall
segments from different sectors of the replica space $j$ and $k$.}
\end{figure}
\end{document}